\documentclass[conference]{IEEEtran}
\IEEEoverridecommandlockouts

\usepackage{cite,comment,url}
\usepackage{caption,subcaption}
\usepackage{amsmath,amssymb,amsfonts}
\usepackage{algorithmic}
\usepackage{graphicx}
\usepackage{textcomp}
\usepackage{xcolor}
\def\BibTeX{{\rm B\kern-.05em{\sc i\kern-.025em b}\kern-.08em
    T\kern-.1667em\lower.7ex\hbox{E}\kern-.125emX}}
\begin{document}

\title{Development of a Metaverse Platform for Tourism Promotion in Apulia\\
\thanks{This work was funded by the TRACECOOP Research project (Italy)   (B96G21000060005).}
}

\author{\IEEEauthorblockN{Enrico Carmine Ciliberti}
\IEEEauthorblockA{\textit{Dep. of Electr., Inf. and Bioeng.} \\
\textit{Polytechnic University of Milan}\\
Milan, Italy \\
enricocarmine.ciliberti@gmail.com}
\and
\IEEEauthorblockN{Marco Fiore}
\IEEEauthorblockA{\textit{Dep. of Electrical and Information Eng.} \\
\textit{Polytechnic University of Bari}\\
Bari, Italy \\
marco.fiore@poliba.it}
\and
\IEEEauthorblockN{Marina Mongiello}
\IEEEauthorblockA{\textit{Dep. of Electrical and Information Eng.} \\
\textit{Polytechnic University of Bari}\\
Bari, Italy \\
marina.mongiello@poliba.it}
}

\maketitle

\begin{abstract}
Metaverse is an engaging way to recreate in a digital environment the real world. It allows people to connect not by just browsing a website, but by using headsets and virtual reality techniques. The metaverse is actually in a rapid development phase, thanks to the advances in different topics. This paper proposes a smart tourism platform in which tourists can interact with guides and different kinds of suppliers, without the need to phisically visit the city they are in. We propose some techniques to scan the real world and transpose it in a metaverse platform, using the recreation of an Italian city, Bari, as a real life scenario.
\end{abstract}


\section{Introduction}
\label{sec:introduction}
The metaverse is frequently characterized as the internet's successor, in which users can interact with each other and with digital objects in three dimensions rather than simply browsing websites or using social media platforms. It provides a variety of new opportunities for entertainment, social interaction, education, and commerce, thank to its direct effect on users satisfaction \cite{suanpang2022extensible}.
Between the different applications of metaverse, we want to focus on smart tourism.\\

Authors of paper \cite{gursoy2022metaverse} raise three main questions with respect to the application of metaverse in tourism area: what could be the staging experiences in the metaverse, how consumers behavior will change, and what business strategies can be developed in this approach. A metaverse tourism ecosystem is defined in \cite{koo2022metaverse}: travelers and suppliers are connected in both the digital and the physical world. The metaverse can provide mirror worlds to virtualize real life experiences. The technology most linked to tourism and metaverse topic is Mixed Reality (MR) \cite{buhalis2022mixed}, in particular in the topic of visiting cultural heritage. MR helps Generation Z people to feel more involved in tourism, creating more engaging adventures from industries.\\

Our proposal aims to develop a metaverse platform to support tourism in Apulia, Italy. In particular, we take advantage of MR technology to let tourists visit Bari, an Apulian city, discovering cultural places and activities. Suppliers can also join this platform to recreate their activities and sponsor them to the public, creating more engagement in visitors. Finally, touristic guides can use their avatar to easily connect to tourists and let them discover the city in a new and entertaining way.

\begin{figure*}
     \centering
     \begin{subfigure}[b]{0.45\textwidth}
         \centering
         \includegraphics[width=\textwidth]{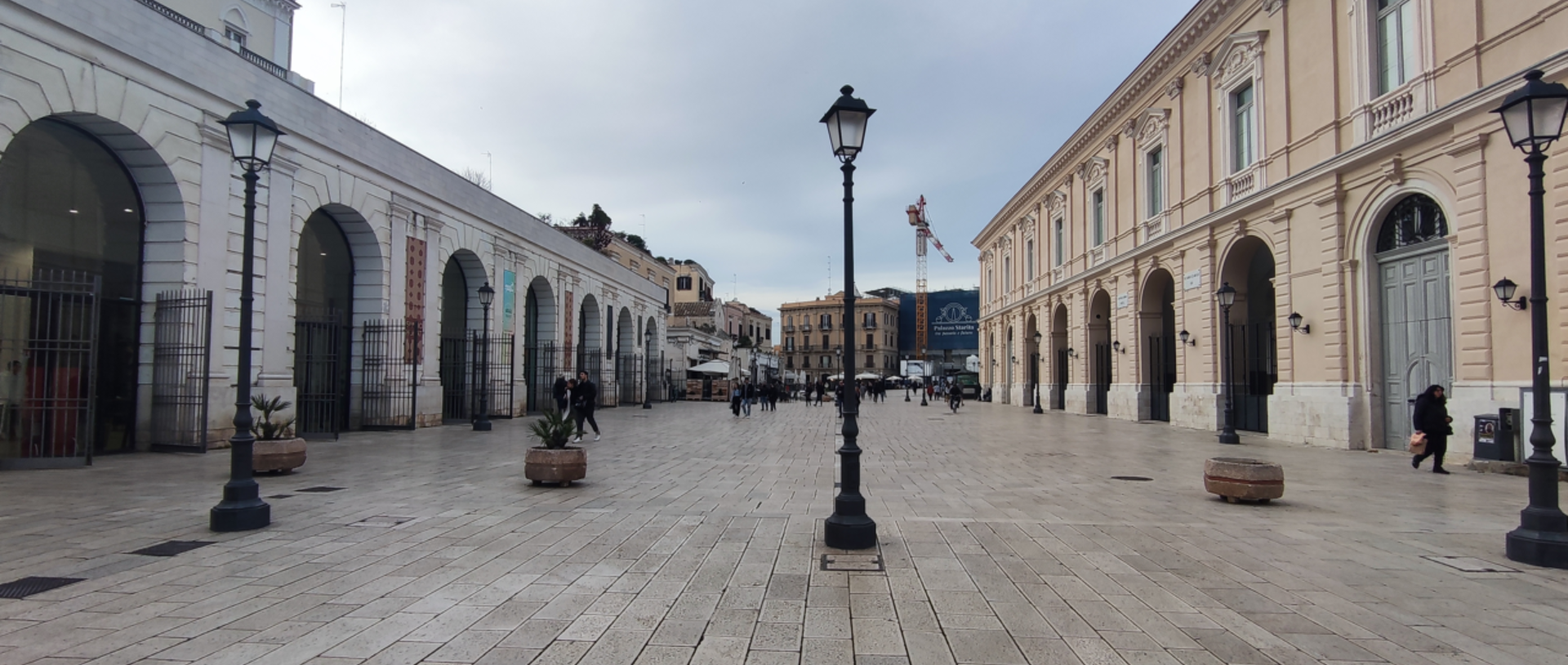}
         \caption{Real photo of Piazza del Ferrarese, Bari, Italy}
         \label{fig:real}
     \end{subfigure}
     \hfill
     \begin{subfigure}[b]{0.45\textwidth}
         \centering
         \includegraphics[width=\textwidth]{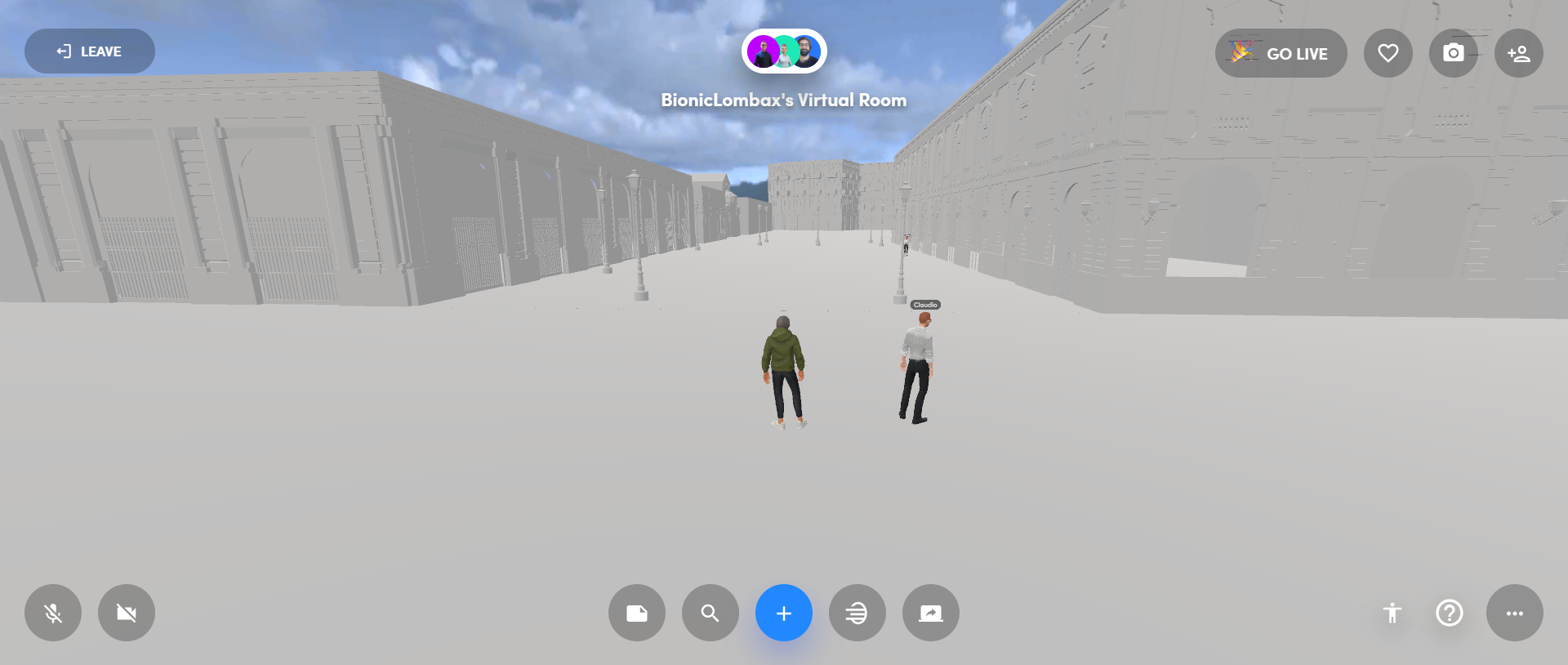}
         \caption{Intermediate phase, without textures}
         \label{fig:intermediate}
     \end{subfigure}
     \hfill
     \begin{subfigure}[b]{0.45\textwidth}
         \centering
         \includegraphics[width=\textwidth]{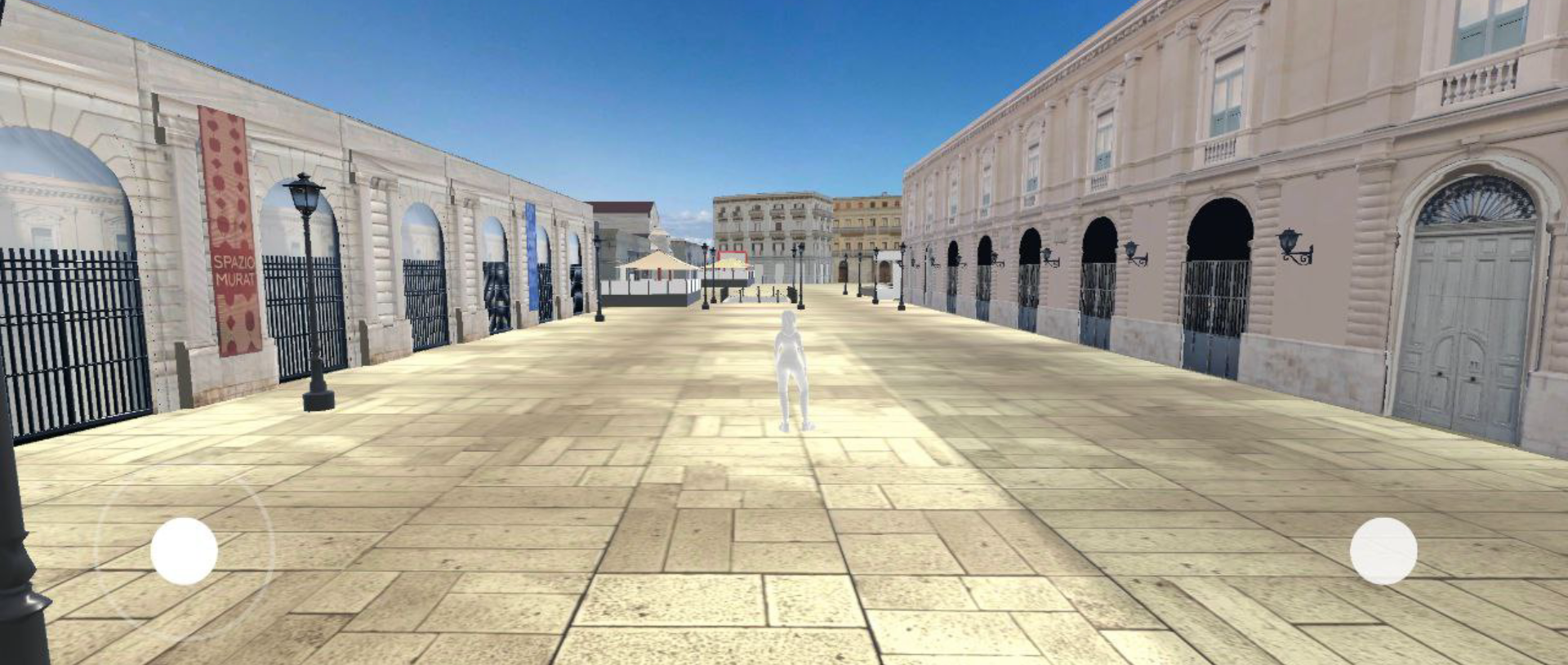}
         \caption{Final phase, with textures and details}
         \label{fig:final}
     \end{subfigure}
     \hfill
     \begin{subfigure}[b]{0.45\textwidth}
         \centering
         \includegraphics[width=\textwidth]{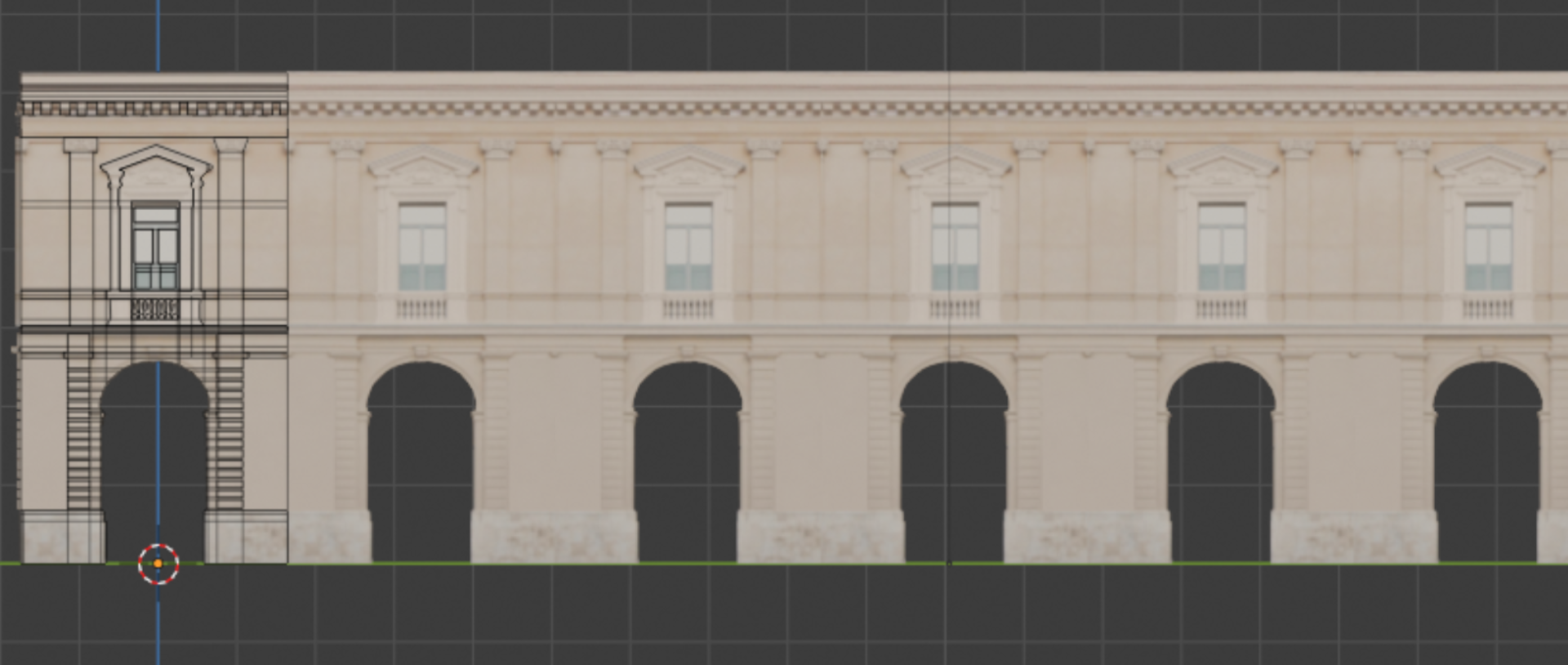}
         \caption{Detail of a building}
         \label{fig:detail}
     \end{subfigure}
        \caption{Development steps for the metaverse platform}
        \label{fig:steps}
\end{figure*}

\section{Our scenario}
\label{sec:scenario}
Our scenario is explained below. The back-end of the architecture is a metaverse platform that allows sharing three-dimensional spaces, usable by different users in real time via devices such as computers, smartphones and virtual reality headsets. The front-end is an immersive virtual space, built with 3D graphics programs and then programmed and loaded into the metaverse. The environment is programmed to possess precise details of real environments, with the addition of extra information elements such as texts, graphics or completely additional 3D objects, available to users under certain conditions to enhance their experience.

Users accessing the platform are either consumers and providers. Consumers are visitors of the platform: they enter the space, attracted by an event inside the platform and then become intrigued by the reproduction of the real space. Furthermore, they have the opportunity to learn and increase their knowledge of the place in an organic way thanks to extra information elements present in the space for advertising, decorative and informative purposes. Providers create the three-dimensional reproduction of the space, and focus on populating the space by organizing cultural themed events in it, promoting the place of interest with exhibitions, historical anecdotes and details interesting geographical locations, thus ensuring the loyalty of passing visitors. Examples of providers are:
\begin{itemize}
    \item \textbf{Touristic guides} who could get in touch with potential tourists and visitors directly from the virtual space, thus following them in detail on any potential question or curiosity.
    \item \textbf{Commercial activities} present in the virtual reproduction of the real world, such as a bar or restaurant, which could have special conventions for those who complete certain challenges inside the space \cite{xu2017serious, pasca2021gamification} or use the advertising space in the classic way.
    \item \textbf{Cultural promotion associations} could use virtual spaces to promote a municipality or locality far beyond its territorial borders, easily reaching the international scale but without sacrificing real events held in presence. This approach guarantees an hybridization of real activities with virtual ones via VR headsets \cite{pietroni2019experience}.
\end{itemize}

\section{Implementation design}
\label{sec:results}
Various steps of implementation are proposed in Fig. \ref{fig:steps}, from the real photo of Piazza del Ferrarese in Bari (Fig. \ref{fig:real}) to an intermediate representation of the location (Fig. \ref{fig:intermediate}), to the final implementation (Fig. \ref{fig:final}). The manufacturing process consists in taking a series of photographs so that the geometry of the object is centered and as straight as possible. We import the image as a plane in Blender\footnote{\url{https://www.blender.org}} and, with the edge looping technique, we underline the salient reliefs of the photograph which are then extruded. To optimize structures with a repeated pattern, an array is used as shown in Fig. \ref{fig:detail}. Preliminary results show a complete platform in the metaverse to explore with or without an headset, thanks to the characteristics of the chosen back-end platform, Spatial\footnote{\url{https://www.spatial.io}}.

\section{Conclusion}
\label{sec:conclusion}
The proposed project will bring a great impact to the tourism market, which has always been looking for new innovative ways to attract the attention of millions of potential tourists every day via web-based social platforms. The main issue with such approaches was the lack of social and human aspects, such as the sociability between people and the sense of discovery.
Users who discover a place through a virtual experience, whether it is actually existing or not, can reach a level of immersion that generates true memories equal to those of a visit to a real place.
Therefore it is of vital importance to exploit this very high level of immersion to create innovative experiences that can amplify the possibilities of tourism based on web3.
Furthermore, this experience has the additional element of sociality, which allows people in the same space to meet and make friends, chat and therefore encourage networking.

Future steps regard the development of additional locations in the platform and an improvement of the quality of textures: a baking of the textures is necessary in order to lighten the overall scene in the best possible way. The overall weight of a Spatial scene is 100 MB. The constant optimization of the manufacturing processes is therefore necessary and auxiliary throughout the course of the creation of the model.

\bibliographystyle{ieeetr}
\bibliography{refs}

\end{document}